\documentclass[11pt,fleqn,twoside]{article}
\usepackage{latexsym}
\makeatletter
\newcommand{\prava}{\footnotesize\it
\begin{flushright}
\begin{minipage}{6cm}
Copyright \copyright 1998 by D.C. Sen and A.R. Chowdhury
\end{minipage}
\end{flushright}}

\newcommand{\name}[1]{\begin{flushleft}
                       \LARGE \bf #1
                       \end{flushleft}\vspace{-3mm}}

\newcommand{\Author}[1]{\begin{flushleft}
                       \it #1 \end{flushleft}}

\newcommand{\Adress}[1]{\begin{flushleft}
                       \it #1 \end{flushleft}}

\newcommand{\Date}[1]{\begin{flushleft}
                      \small  \it #1 \end{flushleft}}

\newcommand{\ehkol}{Author \ name}
\newcommand{\ohkol}{Article \ name}
\renewcommand{\@evenhead}{
\hspace*{-3pt}\raisebox{-15pt}[\headheight][0pt]{\vbox{\hbox to \textwidth
{\thepage \hfil \ehkol}\vskip4pt \hrule}}}
\renewcommand{\@oddhead}{
\hspace*{-3pt}\raisebox{-15pt}[\headheight][0pt]{\vbox{\hbox to \textwidth
{\ohkol \hfil \thepage}\vskip4pt\hrule}}}
\renewcommand{\@evenfoot}{}
\renewcommand{\@oddfoot}{}

\newcommand{\be}{\begin{equation}}
\newcommand{\ee}{\end{equation}}
\newcommand{\ba}{\hspace*{-5pt}\begin{array}}
\newcommand{\ea}{\end{array}}
\newcommand{\p}{\partial}
\newcommand{\ds}{\displaystyle}
\makeatother

\begin{document}
\setcounter{page}{1}

\thispagestyle{empty}

\renewcommand{\ehkol}{D.C. Sen and A.R. Chowdhury}
\renewcommand{\ohkol}{Moyal Quantized BKP Type Hierarchies}

\thispagestyle{empty}

\begin{flushleft}
\footnotesize {\sf
Journal of Nonlinear Mathematical Physics \qquad 1998, V.5, N~1},\ 
\pageref{sen-fp}--\pageref{sen-lp}. \hfill{{\sc Letter}}
\end{flushleft}

\vspace{-5mm}

{\renewcommand{\footnoterule}{}
{\renewcommand{\thefootnote}{}  \footnote{\prava}}

\name{On the Moyal Quantized BKP Type Hierarchies}\label{sen-fp}

\Author{Dolan Chapa SEN and A. Roy CHOWDHURY}

\Adress{High Energy Physics Division, Department of Physics Jadavpur
University,\\ Calcutta 700 032, India}

\Date{Received June 30, 1997; Accepted July 17, 1997}

\begin{abstract}
\noindent
Quantization of BKP type equations are done through the Moyal bracket
and the formalism of pseudo-dif\/ferential operators. It is shown
that a variant of the dressing operator can also be constructed for such
quantized systems.
\end{abstract}

\strut\hfill

\noindent
Quantization of integrable system is one of the most important aspect
of present day research on nonlinear systems.
In two dimensions, a well established methodology was suggested by
Faddeev and his collaborators [1], which goes by the name of Quntum
Inverse Spectral Transform (QISM) [1]. On the other hand, no such
technique is known (which utililes the inverse scattering framework)
for integrable systems in three dimensions. Recently the formalism of
pseudo-dif\/ferential operators was used extensively to study such
systems in 3 (three) dimensions [2] and an ef\/fective way was
found to study the Bi-Hamiltonian structures of such three
dimensional systems. An ingeneous way to derive the quantum version
for the KP system was suggested by Kupershmidt [3] using the tools of
$p$ pseudo-dif\/ferential algebra in conjuction with the idea of the Moyal
bracket [4]. The basic idea is that the quantization is a
deformation of the classical situation. Here, in this communication, we
study a quantization of the BKP like hierarchy [5] using the
Moyal bracket approach. It is demonstrated that such systems can
possess an infinite number of conservation laws; essential for the
complete integrability. Furthermore, we also show that an extension of
the usual dressing operator [6] is possible even for these quantized
or deformed systems.

A quantization is a rule which assigns to every polynomial in
variables $p$ and $q$, a polynomial in operators $p$ and $q$. The
rule must be a linear map satisfying a few natural properties.

For any two operators $f$ and $g$ depending upon the canonical set of
variables $(p_i, q_i)$, the Moyal quantization dictates that the
product be defined according to the following rule:
\be
f*g =\exp \left[ \varepsilon \left( \frac{\p}{\p p_f} \frac{\p}{\p
q_g} -\frac{\p }{\p q_f} \frac{\p}{\p p_g}\right)\right](fg),
\ee
where the operator:
\be
v= \frac{\p}{\p p_f} \frac{\p}{\p
q_g} -\frac{\p }{\p q_f} \frac{\p}{\p p_g}
\ee
acts according to the rule
\be
v(fg)= \frac{\p f}{\p p} \frac{\p g}{\p
q} -\frac{\p f}{\p q} \frac{\p}{\p p}.
\ee
For our situtation, $f$ and $g$ are some pseudo-dif\/ferential
operators. Before proceeding to the actual problem of the quantized
BKP equation, we give below some algebraic formulae which will be
useful later.

}

Using equation (1) we get
\be
\ba{l}
\ds p^i* a = \sum\limits_{s=0} \left( \ba{c} i \\ s \ea \right)
\varepsilon^s a^{(s)} p^{i-s},\\[2mm]
\ds a*p^i = \sum\limits_{s=0} \left( \ba{c} i \\ s \ea \right)
(-\varepsilon)^s a^{(s)} p^{i-s},\\[2mm]
\ds b p^i* a = \sum\limits_{s=0} \left( \ba{c} i \\ s \ea \right)
\varepsilon^s a^{(s)} b p^{i-s},\\[2mm]
\ds p^i* a p^m= \sum\limits_{s=0} \left( \ba{c} i \\ s \ea \right)
\varepsilon^s a^{(s)} p^{i+m-s},\\[2mm]
\ds ap^m*p^i = \sum\limits_{s=0} \left( \ba{c} i \\ s \ea \right)
(-\varepsilon)^s a^{(s)} p^{i+m-s},\\[2mm]
\ds bp^i* ap^m = \sum\limits_{s,k=0}
\frac{\varepsilon^s}{s}\left( \ba{c} s \\ k \ea \right)
\left( \ba{c} i \\ s-k \ea \right)(s-k) \left( \ba{c} m \\ k \ea
\right)k (-1)^k b^{(k)} a^{(s-k)} p^{i+m-s}.
\ea
\ee
Here
$a^{(k)}= \p^k a$. etc., $a$, $b$ are any two arbitrary functions of
$q$ only, and $\p^k$ represents $k$-th order derivative with
respect to $q$, that is $\p^k/\p q^k$. The Moyal commutator is
defined as:
\be
[f,g] =f*g-g*f.
\ee
For example, it can be easily seen that equation (4) leads to
\be
\ba{l}
\ds [p^i, a] =2 \sum\limits_{k=0} \left( \ba{c} i \\ 2k+1 \ea \right)
a^{2k+1}a^{(2k+1)} p^{i-2k-1},\\[2mm]
\ds [bp^i, a] =2 \sum\limits_{k=0} \left( \ba{c} i \\ 2k+1 \ea \right)
\varepsilon^{2k+1}a^{(2k+1)}b p^{i-2k-1},
\ea
\ee
and similarly for others.

In the above expressions $\left( \ba{c} c \\ d \ea \right)$ denotes a
Binomial coef\/ficient with the following usefull properties:
\be
\left( \ba{c} -i-1\\
k_0\ea \right)=
\left\{
\ba{l}
\left( \ba{c} i+2k\\
2k\ea \right), \qquad \mbox{when} \ k_0=2k\\[3mm]
(-1) \left( \ba{c} i+2k+1\\
2k+1\ea \right),
\qquad \mbox{when} \ k_0=2k+1.
\ea
\right.
\ee
For positive values $c$, $d$ these are
\[
\frac{c!}{d!(c-d)!}.
\]

The standart pseudo-dif\/ferential operator $(\overline{\Psi}\mbox{DO})$
is written as:
\be
L=\p+\sum_{m=1} a_m\p^{-m},
\ee
where $a_m$ are functions of ($x$, $y$ and $t$). In the following we
shall rewrite $L$ as
\be
L=p+\sum_{m=1} a_m p^{-m}.
\ee
The deformed or quantized hierarchy of equations are to be obtained from
the Lax equation:
\be
L_t=[P_+,L]=P_+*L-L*P_+.
\ee
Here $P_+=[L^m]_+$, where $m$ is any positive integer and $(+)$ denotes
the only terms with positive powers to be retained. One must also
note that
\be
L^m=L*L*L*L*\cdots *L \qquad (m \ \mbox{times}).
\ee

For the construction of integrable systems, the time flows can be
easily build up with the help of equation (11). For example,
\be
\ba{l}
\ds L^{*p} =\left( p+\sum\limits_{m=1} a_mp^{-m}\right)
*\left(p+\sum_{l=1} a_lp^{-l}\right)\\[2mm]
\ds \phantom{L^{*p}}=p^2+2\sum\limits_ma_mp^{-m+1}+
\sum\limits_{s,k}\varepsilon^s(-1)^k\left(\ba{c} -m\\
s-k \ea \right) \left(\ba{c} -i\\ k\ea\right) a_m^{(k)}
a_i^{(s-k)} p^{-m-i-s}
\ea
\ee
or
\be
\ba{l}
L^{*2} =p^2+2a_1+2a_2p^{-1}+\cdots\\[2mm]
L^{*3}
=p^3+3a_1p+3a_2+(3a_3+3a_1^2+\varepsilon^2a_1^{(2)})p^{-1}+\cdots\\[2mm]
L^{*4}=p^4+4a_1p^2+4a_2p+(4a_3+6a_1^2+\varepsilon^24a_1^{(2)})\\[1mm]
\phantom{L^{*4}=p^4+4a_1p^2+4a_2p}
+(4a_4+12a_1a_2+4\varepsilon^2a_2^{(2)})p^{-1}+\cdots\\[2mm]
L^{*5}=p^5+5a_1p^3+5a_2p^2+(5a_3+10a_1^2+10\varepsilon^2a_2^{(2)})p
\\[1mm]
\phantom{L^{*5}=}
+(5a_4+20a_1a_2+10\varepsilon^2 a_2^{(2)})\\[1mm]
\phantom{L^{*5}=}+\Bigl[
5a_5+20a_1a_3+10a_2^2+10a_1^3+\varepsilon(8a_4^{(1)}+24a_1a_2^{(1)}
+24a_2a_1^{(1)}) \\[1mm]
\phantom{L^{*5}=}+\varepsilon^2(10a_3^{(2)}
+20a_1a_1^{(2)}+10a_1^{(1)} a_1^{(1)})+\varepsilon^3 8a_2^{(3)}+
\varepsilon^4a_1^{(4)}\Bigr]p^{-1}+\cdots\,.
\ea
\ee
The quantized BKP system is then defined through the Lax equations
\be
\frac{\p L}{\p y} = [L_+^3, L]_*\, , \qquad
\frac{\p L}{\p t}=[L_+^5, L]_*.
\ee
Using the above expressions we immediately get:
\be
\ba{l}
a_{1y} =\varepsilon(6a_3^{(1)}+12
a_1a_1^{(1)})+2\varepsilon^3a_1^{(3)}, \\[2mm]
a_{2y}=\varepsilon(6a_4^{(1)}+12
a_1^{(1)}a_2+12a_1a_2^{(1)})+2\varepsilon^3a_2^{(3)}, \\[2mm]
a_{3y}= \varepsilon(6a_5^{(1)}+6a_1a_3^{(3)}+18a_1^{(1)}a_3+12a_2
a_2^{(1)})\\[1mm]
\phantom{a_{3y}=}+
\varepsilon^3(2a_3^{(3)}+6a_1^{(3)}a_1+6a_1^{(2)}a_1^{(1)}),\\[2mm]
a_{4y}=
(6a_6^{(1)}+6a_1a_4^{(1)}+24a_1^{(1)}a_4+18a_2^{(1)}a_3)\\[1mm]
\phantom{a_{4y}=}+\varepsilon^3(2a_4^{(3)}+18 a_1^{(2)}a_2^{(1)}+
24a_1^{(3)} a_2+6a_1a_2^{(3)}).
\ea
\ee
Similar equations for time evolution. Actually, (14) leads to an
infinite set of coupled nonlinear equations in 3 dimensions. To
restrict it to a BKP hierarchy we impose the constraint that
$(L^*)_+^{2n+1}$ will not contain any constant level term. Whence we
get $a_2=a_4=0$, along with
\be
a_{3y}=\varepsilon (6a_5^{(1)} +6a_1 a_3^{(1)} +18a_1^{(1)}a_3)+
\varepsilon^3(2a_3^{(3)}+6a_1^{(3)}a_1+6a_1^{(2)}a_1^{(1)}),
\ee
\be
\ba{l}
a_{1t}=2\varepsilon^5a_1^{(5)}
+\varepsilon^3(20a_3^{(3)}+40a_1^{(3)}a_1 +80a_1^{(2)}a_1^{(1)}
\\[1mm]
\phantom{a_{1t}=}+\varepsilon(10a_5^{(1)}+40a_3^{(1)}a_1+40a_3a_1^{(1)}+
60a_1^2a_1^{(1)}).
\ea
\ee
By eliminating $a_3$ and $a_5$ we obtain
\be
\ba{l}
\ds a_{1t}=-\frac{32}{9} \varepsilon^5 a_1^{(5)} +\varepsilon^3
\left[ -\frac{40}{3} a_1^{(3)}a_1 -\frac{100}{3} a_1^{(1)} a_1^{(2)}
\right] +\varepsilon^2 \frac{20}{9} a_{1y}^{(2)} \\[3mm]
\ds \phantom{a_{1t}=}-10\varepsilon a_1^{2} a_1^{(1)}+
\frac 53(a_1a_{1y} +a_1^{(1)} \p^{-1}a_{1y})+
\frac{5}{18\varepsilon} \p^{-1} a_{1y},
\ea
\ee
which is the required nonlinear equation in (2+1)-dimensions for
$a_1(xyt)$. One can observe that this equation is also fifth order in
derivative with respect to $x$ as in the case of the usual BKP system.

It is possible to generate other types of nonlinear system from
dif\/ferent choices of $y$ and $t$ flows in equation (14). For example,
consider
\be
\ba{l}
\ds \frac{\p L}{\p y}= [L^{*2},L]_*,\\[4mm]
\ds \frac{\p L}{\p t}= [L^{*5},L]_*,
\ea
\ee
where equations (15) are modified to:
\be
\ba{l}
a_{1y}=4\varepsilon a_2^{(1)},\\
a_{2y} =4\varepsilon a_3^{(1)} +4\varepsilon a_1 a_1^{(1)},\\
a_{3y}=4\varepsilon a_4^{(1)} +8\varepsilon a_1^{(1)} a_2,\\
a_{4y}= 4\varepsilon a_5^{(1)} +12 \varepsilon a_1^{(1)}a_3
+4\varepsilon^3a_1a_1^{(3)}
\ea
\ee
along with
\be
\ba{l}
a_{1t}= 2\varepsilon^5 a_1^{(5)} +\varepsilon^3 (20 a_3^{(3)}+
40a_1^{(3)}a_1 +80 a_1^{(2)} a_1^{(1)}) \\[2mm]
\phantom{a_{1t}=}+\varepsilon(10a_5^{(1)} +40a_3^{(1)}a_1
+30a_3a_1^{(1)}+ 40 a_2a_2^{(1)} +60a_1^2a_1^{(1)}).
\ea
\ee
It is interesting to note that one can eliminate all the variables
other than $a_1$, without imposing any extra condition on
$L_+^{2n+1}$, since equation (20) yields:
\be
\ba{l}
\ds a_2=\frac{1}{4\varepsilon} \p^{-1} a_{1y},
\\[3mm]
\ds a_3= \frac{1}{16 \varepsilon^2} \p^{-2} a_{1yy}-\frac 12 a_1^2,
\\[3mm]
\ds a_4 = \frac{1}{4\varepsilon} \p^{-1} a_{3y}
-2\p^{-1}(a_1^{(1)}a_2).
\ea
\ee
Thus, from (21) we at once obtain
\be
\ba{l}
\ds a_{1t}=2\varepsilon^5 a_1^{(5)} +\varepsilon^3\left[
10a_1 a_1^{(3)} +20a_1^{(1)} a_1^{(2)}\right]+
\varepsilon\left[ 15a_1^2 a_1^{(1)} +\frac{5}{4}
a_{1yy}^{(1)}\right]
\\[3mm]
\hspace*{-2pt}\phantom{a_{1t}=}\ds +\frac{5}{8\varepsilon}
\left[3\p^{-1}(a_{1y}a_{1y} +a_{1y}a_{1yy})+2a_{1y}\p^{-1} a_{1y}
-3a_1^{(1)} \p^{-2} a_{1yy} +4a_1\p^{-1}a_{1yy}\right]\\[3mm]
\phantom{a_{1t}=}
\ds -\frac{5}{128\varepsilon^3} \p^{-3}a_{14y}.
\ea
\ee
It each case, as in the case of usual $\overline{\Psi}\mbox{DO}$
approach, the conserved quantities are obtained as
$\mbox{Res.}(L^{2n+1})$, where $\mbox{Res}$ denote the coef\/ficient
of $p^{-1}$. We have checked that such residues turn out to be
combinations of total derivatives with respect to $(x, y, t)$.

We now report on intriguing fact about such Moyal quantized systems.
This is related to the so-called modified system. In an interesting
paper it was observed by Kupershmidt that modified equations can be
generated if one uses
\be
\frac{d L}{d t} =[(L^m)_{\geq 1}, L]
\ee
instead of $(L^m)_{\geq 0}$. (In some cases one can also consider
$(L^m)_{\geq 2}$.) Here we observe that, if we consider
\be
\frac{\p L}{\p y} =[L^3_{\geq 1}, L]_*\ , \qquad
\frac{\p L}{\p t} =[L^5_{\geq 1}, L]_*\ ,
\ee
and using
\be
\ba{l}
(L^3)_{\geq 1}=p^3+3a_1p,\\[2mm]
(L^{*5})_{\geq 1}=p^5+5a_1p^3+5a_2p^2+(5a_3+10a_1^2+\varepsilon^2
10a_1^{(2)})p
\ea
\ee
in these equations, we get
\be
\ba{l}
a_2^{(1)}=0,\\
a_{1y}=6\varepsilon a_3^{(1)}+12 \varepsilon a_1 a_1^{(1)}
+2\varepsilon^3a_1^{(3)}, \\
a_{2y}=6\varepsilon a_4^{(1)} +12 \varepsilon a_2 a_1^{(2)},\\
a_{3y}= 6\varepsilon a_5^{(1)} +6\varepsilon a_1a_3^{(1)} +18
\varepsilon a_1^{(1)} a_3 +\varepsilon^3 (2a_3^{(3)}+
6a_1a_1^{(3)} +6a_1^{(1)}a_1^{(2)})
\ea
\ee
alongwith
\[
\ba{l}
a_{1t} =2\varepsilon^5 a_1^{(5)} +\varepsilon^3 (20 a_3^{(3)}+40
a_1a_1^{(3)} +80 a_1^{(1)} a_1^{(2)})\\[2mm]
\phantom{a_{1t}=} +\varepsilon(40 a_1a_3^{(1)}+40 a_1^{(1)} a_3)
+60 a_4^2 a_1^{(1)} +10a_5^{(1)}.
\ea
\]
Again eliminating variables other than $a_1$, we get
\be
\ba{l}
a_{1t} =2\varepsilon^5 a_1^{(5)} +\varepsilon^3 (20 a_3^{(3)}+40
a_1^{(3)}a_1 +80 a_1^{(2)} a_1^{(1)})\\[2mm]
\phantom{a_{1t}=} +\varepsilon(40 a_1a_3^{(1)}+40 a_1^{(1)} a_3
+60 a_1^2 a_1^{(1)} +10a_5^{(1)}).
\ea
\ee
So we get back equation (21), but this time without any restriction on
the constant level term of $(L^*)^{2n+1}$.

From the structure of the nonlinear equations discuseed so far, it
appears that one can define an analogue of a dressing operator even in
the Moyal quantized systems.

Let us set
\be
s=1+\sum_{m=1}^\infty \omega_mp^{-m}
\ee
for the deressing operator. Now 
\be
s^{-1} =1+\sum_{m=1}^\infty u_mp^{-m},
\ee
with $s+s^{-1} =1$, leads to
\be
(u_m +\omega_m) p^{-m} +\varepsilon^s (-1)^k
\left( \ba{c} -m\\ j-k \ea \right)
\left( \ba{c} -j\\ k\ea \right) \omega_m^{(k)} u_j^{(s-k)}
p^{-m-js}=0,
\ee
from which one can express the $u$'s in terms of the $\omega$'s. For
example,
\be
\ba{l}
u_1=-\omega_1,\\
u_2=-\omega_2+\omega_1^2,\\
u_2=-\omega_3-\omega_1^2+2\omega_1\omega_2, \ \ldots, \ \mbox{etc.}
\ea
\ee
Now we demand that
\be
s*p *s^{-1} =L=p+\sum_{m=1} a_mp^{-m}\ ,
\ee
which immediately leads to
\be
\ba{l}
a_1 =-2\varepsilon \omega_1^{(1)},\\
a_2 =\varepsilon(-2 \omega_2^{(1)} +2\omega_1 \omega_1^{(1)}),\\
a_3=u_4 +\omega_4 +\omega_1 u_3 +\omega_2u_2 +\omega_3
u_1 \\
\phantom{a_3=}+\varepsilon (u_3^{(1)} -\omega_3^{(1)}+\omega_1^{(1)}u_2
-\omega_2u_1^{(1)} )+\varepsilon^2 \omega_1^{(1)} u_1^{(1)},
\ \ldots, \ \mbox{etc.}
\ea
\ee
On the other hand, if we compute $s*p^3*s^{-1}$ and $s*p^5*s^{-1}$ via
the rules (4), we get
\be
\ba{l}
\ds s*p^3*s^{-1}= \left(1+\sum\limits_m \omega_m p^{-m}\right)*p^3 *
\left(1+\sum\limits_{l} u_lp^{-l}\right)\\[2mm]
\phantom{s*p^3*s^{-1}}=p^3+(\omega_1+u_1)p^2
+(u_2+\omega_2+\omega_1u_1 +3\varepsilon u_1^{(1)}-3\varepsilon
\omega_1^{(1)})p+\cdots\ .
\ea
\ee
Whence, using equations (32) and (34), we get
\be
\ba{l}
(s*p^3 *s^{-1})_+ =p^3+3a_1p +(3\varepsilon u_2^{(1)} -3\varepsilon
\omega_2^{(1)}) \\
\phantom{(s*p^3 *s^{-1})_+ }=p^3+3a_1p+3a_2\\
\phantom{(s*p^3 *s^{-1})_+ }=L^3 \ \ \mbox{(as given in equation
(13))}.
\ea
\ee

Using the same methodology, but with a more laborious computation, we
can prove that
\be
\ba{l}
(s*p^5*s^{-1})_+ =p^5 +5a_1p^3 +5a_2p^2 +(5a_3+10 a_1^2+10a_1^2
a_1^{(2)})p\\
\phantom{(s*p^5*s^{-1})_+ =} +(5a_4+20a_1a_2+10 \varepsilon^2
a_2^{(2)}).
\ea
\ee

In the above analysis we have shown that nonlinear systems in
(2+1)-dimensions, which are in the category of the BKP equation, can be
quantized using the pseudo-dif\/ferential operators and Moyal bracket
formalism. They are completely integrable since there are an infinite
number of conserved quantities with them.

One of the authors (D.C. Sen) is gratefull to CSIR (Govt. of India)
for suppport though a SRF grant.

\label{sen-lp}
\end{document}